\documentclass[aps,amsmath,reprint]{revtex4-1}

\usepackage{graphicx}

\begin{document}
\title{Computing Bayes factors to measure evidence from experiments: An extension of the BIC approximation}
\author{Thomas J. Faulkenberry}
\email{faulkenberry@tarleton.edu}
\affiliation{Tarleton State University}

\begin{abstract}
  Bayesian inference affords scientists with powerful tools for testing hypotheses.  One of these tools is the Bayes factor, which indexes the extent to which support for one hypothesis over another is updated after seeing the data. Part of the hesitance to adopt this approach may stem from an unfamiliarity with the computational tools necessary for computing Bayes factors. Previous work has shown that closed form approximations of Bayes factors are relatively easy to obtain for between-groups methods, such as an analysis of variance or $t$-test.  In this paper, I extend this approximation to develop a formula for the Bayes factor that directly uses information that is typically reported for ANOVAs (e.g., the $F$ ratio and degrees of freedom).  After giving two examples of its use, I report the results of simulations which show that even with minimal input, this approximate Bayes factor produces similar results to existing software solutions.\\

\noindent
Note: to appear in \textit{Biometrical Letters}.
\end{abstract}

\keywords{Bayes factors, Bayesian inference, analysis of variance, hypothesis testing}

\maketitle

\section{Introduction}

Hypothesis testing is the primary tool for statistical inference across much of the biological and behavioral sciences. As such, most scientists are trained in classical null hypothesis significance testing (NHST). The scenario for testing a hypothesis is likely familiar to most readers of this journal.  Suppose one wants to test a specific research hypothesis (e.g., some treatment has an effect on some outcome measure). NHST works by first assuming a \emph{null hypothesis} (e.g., the treatment has \emph{no effect}) and then computing some test statistic for a sample of data.  This sample test statistic is then compared to a hypothetical distribution of test statistics that would arise if the null hypothesis were true.  If the sample's test statistic is in the tail of the distribution (that is, it should occur with low probability), the scientist decides to \emph{reject} the null hypothesis in favor of the alternative hypothesis.  Further, the $p$-value, which indicates how surprising the sample would be if the null hypothesis were true, is often taken as a measure of evidence: the lower the $p$-value, the \emph{stronger} the evidence.

While orthodox across many disciplines, NHST does have philosophical criticisms \citep{wagenmakers2007}. Also, the $p$-value is prone to misinterpretation \citep{gigerenzer2004,hoekstra2014}.  Finally, NHST is ideally suited to providing support for the alternative hypothesis, but the procedure does not work in the case where one wants to measure support for the null hypothesis. That is, we can reject the null, but we cannot \emph{accept} the null. To overcome this limitation, we can use an alternative method for testing hypotheses that is based on Bayesian inference: the Bayes factor. 

\subsection{The Bayes factor}

Bayesian inference is a method of measurement that is based on the computation of $P(H \mid D)$, which is called the {\it posterior probability} of a hypothesis $H$, given data $D$.  Bayes' theorem casts this probability as

\begin{equation}\label{eq:bayes}
  P(H\mid D) = \frac{P(D\mid H) \cdot P(H)}{P(D)}.
\end{equation}

\noindent
One may think of Equation \ref{eq:bayes} in the following manner: before observing data $D$, one assigns a {\it prior probability} $P(H)$ to hypothesis $H$.  After observing data, one can then update this prior probability to a {\it posterior probability} $P(H\mid D)$ by multiplying the prior $P(H)$ by the likelihood $P(D\mid H)$.  This product is then rescaled to a probability distribution (i.e., total probability = 1) by dividing by the {\it marginal probability} $P(D)$.

Bayes' theorem provides a natural way to test hypotheses. Suppose we have two competing hypotheses: an alternative hypothesis $H_1$ and a null hypothesis $H_0$.  We can directly compare the posterior probabilities of $H_1$ and $H_0$ by computing their ratio; that is, we can compute the {\it posterior odds} in favor of $H_1$ over $H_0$ as $P(H_1\mid D) / P(H_0\mid D)$.  Using Bayes' theorem (Equation \ref{eq:bayes}), it is trivial to see that

\begin{equation}\label{eq:odds}
  \underbrace{\frac{P(H_1\mid D)}{P(H_0\mid D)}}_{\text{posterior odds}} = \underbrace{\frac{P(D\mid H_1)}{P(D\mid H_0)}}_{\text{Bayes factor}} \cdot \underbrace{\frac{P(H_1)}{P(H_0)}}_{\text{prior odds}}.
\end{equation}

This equation can also be interpreted in terms of the ``updating'' metaphor that was explained above.  Specifically, the posterior odds are equal to the prior odds multiplied by an updating factor.  This updating factor is equal to the ratio of likelihoods $P(D\mid H_1)$ and $P(D\mid H_0)$, and is called the {\it Bayes factor} \citep{jeffreys1961}.  Intuitively, the Bayes factor can be interpreted as the {\it weight of evidence} provided by a set of data $D$.  For example, suppose that one assigned the prior odds of $H_1$ and $H_0$ equal to 1; that is, $H_1$ and $H_0$ are {\it a priori} assumed to be equally likely.  Then, suppose that after observing data $D$, the Bayes factor was computed to be 10.  Now, the posterior odds (the odds of $H_1$ over $H_0$ {\it after} observing data) is 10:1 in favor of $H_1$ over $H_0$.  As such, the Bayes factor provides an easily interpretable measure of the evidence in favor of $H_1$. 

In order to help with interpreting Bayes factors, various classification schemes have been proposed.  One simple scheme is a four-way classification proposed by \citet{raftery1995}, where Bayes factors between 1 and 3 are considered {\it weak} evidence; between 3 and 20 constitutes {\it positive} evidence; between 20 and 150 constitutes {\it strong} evidence; and beyond 150 is considered {\it very strong} evidence.

Note that in the discussion above, there was no specific assumption about the order in which we addressed $H_1$ and $H_0$.  If instead we wanted to assess the weight of evidence in favor of $H_0$ over $H_1$, Equation \ref{eq:odds} could simply be adjusted by taking reciprocals.  As such, implied direction is important when computing Bayes factors, so one must be careful to define notation when representing Bayes factors.  A common convention is to define $BF_{10}$ as the Bayes factor for $H_1$ over $H_0$; similarly, $BF_{01}$ would represent the Bayes factor for $H_0$ over $H_1$.  Note that $BF_{10} = 1 / BF_{01}$.

In summary, the Bayes factor provides an index of preference for one hypothesis over another that has some advantages over NHST.  First, the Bayes factor tells us by how much a data sample should update our belief in one hypothesis over a competing one.  Second, though NHST does not allow one to accept a null hypothesis, doing so within a Bayesian framework makes perfect sense.  Given these advantages, it may be surprising that Bayesian inference has not been used more often in the empirical sciences. One reason for the lack of more widespread adoption may be that Bayes factors are quite difficult to compute.  We tackle this issue in the next section.

\section{Computing Bayes factors}

As an example, suppose we are interested in computing the Bayes factor for a null hypothesis $H_0$ over an alternative hypothesis $H_1$, given data $D$.  Recall from Equation \ref{eq:odds} that this Bayes factor (denoted $BF_{01}$) is equal to

\[
  BF_{01} = \frac{P(D\mid H_0)}{P(D\mid H_1)}.
\]

\noindent
While this equation may seem conceptually quite simple, it is computationally much more difficult.  This is because in order to compute the numerator and denominator, one must parameterize the hypotheses (or {\it models}, to be more clear), and then each likelihood is computed by conditioning over all possible parameter values and summing over this set.  Since these potential parameter values are often over a continuous parameter space, this computation requires integration, and thus the formula for the Bayes factor amounts to

\begin{equation}\label{eq:integral}
  BF_{01} = \frac{\int_{\theta \in \Theta_0} P(D\mid H_0,\theta) \pi_0(\theta) d\theta}{\int_{\theta \in \Theta_1} P(D\mid H_1,\theta) \pi_1(\theta) d\theta}
\end{equation}

\noindent
where $\Theta_0$ and $\Theta_1$ are the parameter spaces for models $H_0$ and $H_1$, respectively, and $\pi_0$ and $\pi_1$ are the prior probability density functions of the parameters of $H_0$ and $H_1$, respectively.

Thus, in order to compute $BF_{01}$, one must specify the priors $\pi_0$ and $\pi_1$ for $H_0$ and $H_1$.  Further, the integrals usually do not have closed-form solutions, so numerical integration techniques are necessary.  These requirements lend a computation of the Bayes factor to be inaccessible to all but the most ardent researchers who have at least a more-than-modest amount of mathematical training.

Fortunately, there are an increasing number of solutions that avoid a direct encounter with computations of the above type.  Recently, researchers have proposed {\it default} priors for standard experimental designs such as $t$-tests \citep{rouder2009,morey2011} and ANOVA \citep{rouder2012}.  These default priors are implemented in software packages such as the R package BayesFactor \citep{bayesfactor}, and as such, have provided a user-friendly method for researchers to compute Bayes factors without the computational overhead needed in Equation \ref{eq:integral}. While these software solutions work quite well for computing Bayes factors from raw data, they are a bit limited in the following context.  Suppose that in the course of reading some published literature, a researcher comes across a result that is presented as ``nonsignificant'', with associated test statistic $F(1,23)=2.21$, $p=0.15$.  In an NHST context, this nonsignificant result does not provide evidence \emph{for} the null hypothesis; rather, it just implies that we \emph{cannot reject} the null. A natural question would be what, if any, support does this result provide for the null hypothesis?  Of course, a Bayes factor would be useful here, but without the raw data, we cannot use the previously mentioned software solutions. To this end, it would be advantageous if there were some easy way to compute a Bayes factor directly from the reported test statistic.

It turns out that this computation is indeed possible, at least in certain cases.  In the following, I will show how one particular method for computing Bayes factors \citep[the BIC approximation;][]{raftery1995} can be adapted to solve this problem, thus allowing researchers to compute approximate Bayes factors from summary statistics alone (with no need for raw data).  Further, I will show through simulations that this method compares well to the default Bayes factors for ANOVA developed by \citet{rouder2012}.

\section{The BIC approximation of the Bayes factor}

\citet{wagenmakers2007} demonstrated a method \citep[based on earlier work by][]{raftery1995} for computing approximate Bayes factors using the BIC (Bayesian Information Criterion).  For a given model $H_i$, the BIC is defined as

\[
  \text{BIC}(H_i) = -2\log L_i + k_i\cdot \log n,
\]

\noindent
where $n$ is the number of observations, $k_i$ is the number of free parameters
of model $H_i$, and $L_i$ is the maximum likelihood for model $H_i$.  He then showed that the Bayes factor for $H_o$ over $H_1$ can be approximated as

\begin{equation}
  \label{eq:BF}
  BF_{01} \approx \exp\Bigl(\Delta\text{BIC}_{10}/2\Bigl),
\end{equation}

\noindent
where $\Delta\text{BIC}_{10} = \text{BIC}(H_1)-\text{BIC}(H_0)$.  Further, \citet{wagenmakers2007} showed that when comparing an alternative hypothesis $H_1$ to a null hypothesis $H_0$, 

\begin{equation}\label{eq:BIC}
  \Delta\text{BIC}_{10} = n\log\Biggl(\frac{SSE_1}{SSE_0}\Biggr) + (k_1-k_0)\log n. 
\end{equation}

\noindent
In this equation, $SSE_0$ and $SSE_1$ represent the sum of squares for the error terms in models $H_0$ and $H_1$, respectively.  Both \citet{wagenmakers2007} and \citet{masson2011} give excellent examples of how to use this approximation to compute Bayes factors, assuming one is given information about $SSE_0$ and $SSE_1$, as is the case with most statistical software.  However, we will now consider the situation where one is given the statistical summary (i.e., $F(1,23)=2.21$), but not the ANOVA output.

Suppose we wish to examine an effect of some independent variable with associated $F$-ratio $F(df_1,df_2)$, where $df_1$ represents the degrees of freedom associated with the manipulation, and $df_2$ represents the degrees of freedom associated with the error term.  Then, $F = \frac{SS_1/df_1}{SS_2/df_2} = \frac{SS_1}{SS_2}\cdot \frac{df_2}{df_1}$, where $SS_1$ and $SS_2$ are the sum of squared errors associated with the manipulation and the error term, respectively. 

From Equation \ref{eq:BIC}, we see that

\begin{align*}
  \Delta\text{BIC}_{10} &= n\log\left(\frac{SSE_1}{SSE_0}\right) + (k_1-k_0)\log n\\
                        & = n\log \left(\frac{SS_2}{SS_1+SS_2}\right) + df_1\log n.
\end{align*}

\noindent
This equality holds because $SSE_1$ represents the sum of squares that is not explained by $H_1$, which is simply $SS_2$ (the error term).  Similarly, $SSE_0$ is the sum of squares not explained by $H_0$, which is the sum of $SS_1$ and $SS_2$ \citep[see][p. 799]{wagenmakers2007}.  Finally, in the context of comparing $H_1$ and $H_0$ in an ANOVA design, we have $k_1-k_0=df_1$.  Now, we can use algebra to re-express $\Delta\text{BIC}_{10}$ in terms of $F$:

\begin{align*}
  \Delta\text{BIC}_{10} &=  n\log \left(\frac{SS_2}{SS_1+SS_2}\right) + df_1\log n\\
  &= n\log\left(\frac{1}{\frac{SS_1}{SS_2}+1}\right) + df_1\log n\\
                        &= n\log \left( \frac{\frac{df_2}{df_1}}{\frac{SS_1}{SS_2}\cdot \frac{df_2}{df_1}+\frac{df_2}{df_1}}\right) + df_1\log n\\
                        &= n\log \left(\frac{\frac{df_2}{df_1}}{F + \frac{df_2}{df_1}}\right) + df_1\log n\\
  & = n\log\left(\frac{df_2}{Fdf_1 + df_2}\right) + df_1\log n.\\
\end{align*}

\noindent
Substituting this into Equation \ref{eq:BF}, we can compute:
\begin{align*}
  BF_{01} & \approx \exp\left(\Delta\text{BIC}_{10}/2\right)\\
  & = \exp\left[ \frac{1}{2} \left( n\log\left(\frac{df_2}{Fdf_1 + df_2}\right) + df_1\log n\right)\right]\\
          & = \exp \left[\frac{n}{2}\log\left(\frac{df_2}{Fdf_1+df_2}\right) + \frac{df_1}{2}\log n\right]\\
          & = \left(\frac{df_2}{Fdf_1+df_2}\right)^{n/2} \cdot n^{df_1/2}\\
          & = \sqrt{\frac{df_2^n \cdot n^{df_1}}{(Fdf_1+df_2)^n}}\\
          &= \sqrt{\frac{n^{df_1}}{\left(\frac{Fdf_1}{df_2}+1\right)^n}}.
\end{align*}

Rearranging this last expression slightly yields the approximation:
\begin{equation}\label{eq:BIC2}
BF_{01} \approx \sqrt{n^{df_1}\left(1+\frac{Fdf_1}{df_2}\right)^{-n}}
\end{equation}

Practically speaking, the approximation given in Equation \ref{eq:BIC2} offers nothing new over the previous formulations of the BIC approximation given in \citet{wagenmakers2007} and \citet{masson2011}.  However, it does have two advantages over these previous formulations.  First, one can directly take reported ANOVA statistics (e.g., sample size, degrees of freedom, and the $F$-ratio) and compute $BF_{01}$ without having to compute $SSE_0$ or $SSE_1$.  We should note that \citet{masson2011} correctly mentions that $SSE_1/SSE_0 = 1-\eta_p^2$, so if a paper reports $\eta_p^2$, the need for computing $SSE_0$ and $SSE_1$ is nullified.  However, the method of \citet{masson2011} is still essentially a two-step process; one first computes $\Delta\text{BIC}_{10}$, which in turn is used to compute $BF_{01}$.  In contrast, the expression derived in Equation \ref{eq:BIC2} is a one-step process that can easily be implemented using a scientific calculator or a simple spreadsheet.

\section{Example computations}

In this section, we will discuss two examples of using Equation \ref{eq:BIC2} to compute Bayes factors. In the first example, I will show how to compute and interpret a Bayes factor for a reported null effect in the field of experimental psychology. In the second example, I will show how to modify Equation \ref{eq:BIC2} to work with an independent samples $t$-test.

\subsection{Example 1}

\citet{sevos2016} performed an experiment to assess whether schizophrenics could internal simulate motor actions when perceiving graspable objects. The evidence for such internal simulation comes from a statistical interaction between response-orientation compatibility and the presence of an individual name prime.  Sevos et al. reported that in a sample of $n=18$ schizophrenics, there was no interaction between this compatibility and name prime, $F(1,17)=2.584$, $p=0.126$.  Critically, Sevos et al. claimed that this null effect was evidence for the absence of sensorimotor simulation, which, as they indicate, would imply that schizophrenics would have to rely on higher cognitive processes for even the most simple daily tasks.  This claim is based on a null effect, which as pointed out earlier, is problematic for a null hypothesis testing framework.  I will now show how to compute a Bayes factor $BF_{01}$ to assess the evidence for this null effect.

To this end, we use Equation \ref{eq:BIC2} to compute

\begin{align*}
BF_{01} &\approx \sqrt{n^{df_1}\left(1+\frac{Fdf_1}{df_2}\right)^{-n}}\\
        & = \sqrt{18^1\left(1+\frac{2.584\cdot 1}{17}\right)^{-18}}\\
  &= 1.187.
\end{align*}

This Bayes factor can be interpeted as follows: after seeing the data, our belief in the null hypothesis should increase only by a factor of 1.19.  In other words, this data is not very informative toward our belief in the null, which implies that the claim of a null effect in \citet{sevos2016} may be a bit optimistic.  According the classification scheme of \citet{raftery1995}, this result provides \emph{weak} evidence for the null.

\subsection{Example 2}

\citet{borota2014} observed that with a sample of $n=73$ participants, those who received 200 mg of caffeine performed significantly better on a test of object memory compared to a control group of participants who received a placebo, $t(71)=2.0$, $p=0.049$. \citet{borota2014} claimed this result as evidence that caffeine enhances memory consolidation.

As before, we can measure the evidence provided from this data sample by computing a Bayes factor.  However, we note that because Equation \ref{eq:BIC2} casts the Bayes factor in terms of an $F$-ratio, it may not be immediately obvious whether we can use Equation \ref{eq:BIC2} in this context.  It turns out to be straightforward to modify Equation \ref{eq:BIC2} to work for an independent samples $t$-test.  All we need are two simple transformations: (1) $F=t^2$, and (2) $df_1=1$. Applying these to Equation \ref{eq:BIC2}, we get

\begin{align*}
BF_{01} &\approx \sqrt{n^{df_1}\left(1+\frac{Fdf_1}{df_2}\right)^{-n}}\\
        &= \sqrt{n^1\left(1+\frac{t^2(1)}{df_2}\right)^{-n}}\\
        &= \sqrt{n\left(1+\frac{t^2}{df_2}\right)^{-n}}\\
\end{align*}

We can now apply this equation to the reported results of \citet{borota2014}.  We see that

\begin{align*}
  BF_{01} &\approx \sqrt{n\left(1+\frac{t^2}{df_2}\right)^{-n}}\\
          &= \sqrt{73\left(1+\frac{(2.0)^2}{71}\right)^{-73}}\\
  &= 1.16\\
\end{align*}

So, perhaps counterintuitively, the significant result reported in \citet{borota2014} turns out to be weak evidence in support of the \emph{null}!  Such results are an example of Lindley's paradox \citep{lindley1957}, where ``significant'' $p$ values between 0.04 and 0.05 can actually imply evidence in favor of the null when analyzed in a Bayesian framework.  

\section{Simulations: BIC approximation versus default Bayesian ANOVA}

At this stage, it is clear that Equation \ref{eq:BIC2} provides a straightforward method for computing an approximate Bayes factor, especially in cases when one is given only minimal output from a reported ANOVA or $t$ test.  However, it is not yet clear to what extent this BIC approximation would result in the same decision if a Bayesian analysis of variance \citep[e.g.,][]{rouder2012} were performed on the raw data.  To answer this question, I performed a series of simulations.

Each simulation consisted of 1000 randomly generated data sets under a $2\times 3$ factorial design. The choice of this design is to replicate experimental conditions that are common across many applications in the biological and behavioral sciences.  Further, I simulated varying common levels of statistical power in these experiments by testing 3 different cell-size conditions: $n=20, 50,$ or $80$. Specifically each data set consisted of a vector $\mathbf{y}$ generated as

\[
  y_{ijk} = \alpha_i + \tau_j + \gamma_{ij} +\varepsilon_{ijk}
\]

\noindent
where $i=1,2$, $j=1,2,3$, and $k=1,\dots, n$.  The ``effects'' $\alpha$, $\tau$, and $\gamma$ were generated from multivariate normal distributions with mean 0 and variance $g$, yielding three different effect sizes obtained by setting $g = 0, 0.05,$ and $0.2$ \citep[as in][]{wang2017}.  In all, there were 9 different simulations, generated by crossing the 3 cell sizes ($n=20, 50, 80$) with the 3 effect sizes ($g=0, 0.05, 0.2$).

For each data set, I computed (1) a Bayesian ANOVA using the BayesFactor package in R \citep{bayesfactor} and (2) the BIC approximation using Equation \ref{eq:BIC2} from the traditional ANOVA.  Bayes factors were computed as $BF_{10}$ to assess evidence in favor of the alternative hypothesis over the null hypothesis.  Similar to \citet{wang2017}, I set the decision criterion to select the alternative hypothesis if $\log(BF)>0$, and the null hypothesis otherwise.  Because the different cell sizes resulted in similar outcomes, for brevity I only report the $n=50$ cell size condition in the summaries below.  Also note that all BayesFactor models were fit with a ``wide'' prior, which is roughly equivalent to the unit-information prior used by \citet{raftery1995} for the BIC approximation.

First, I will report the results of computing Bayes factors for the main effect $\alpha$ in each of the effect size conditions $g=0$, $g=0.05$, and $g=0.2$. Five-number summaries for $\log(BF)$ are reported for the $n=50$ simulation in Table \ref{tab:summary1}, as well as the proportion of simulated data sets for which the Bayesian ANOVA and the BIC approximation from Equation \ref{eq:BIC2} selected the same model.  

\begin{table*}
\caption{\label{tab:summary1}%
Summary of values for $\log BF$ for main effect $\alpha$ with cell size $n=50$}
\begin{ruledtabular}
\begin{tabular}{cccccccc} 
$g$ & $BF$ type & Min & $Q_1$ & Median & $Q_3$ & Max & Consistency\\
\colrule
0 & BayesFactor & -2.40 & -2.35 & -2.14 & -1.74 & 3.38 & \\
& BIC & -2.85 & -2.80 & -2.59 & -2.17 & 3.17 & 0.985 \\[2mm]

0.05 & BayesFactor & -2.40 & -1.83 & -0.20 & 4.24 & 47.49 & \\
& BIC & -2.85 & -2.23 & -0.41 & 4.44 & 53.29 & 0.984\\[2mm]

0.2 & BayesFactor & -2.40 & -0.61 & 4.88 & 17.26 & 119.69 & \\
& BIC & -2.85 & -0.63 & 6.76 & 21.71 & 146.02 & 0.987\\
\end{tabular}
\end{ruledtabular}
\end{table*}

\begin{figure*}
  \centering
  \includegraphics[width=\textwidth]{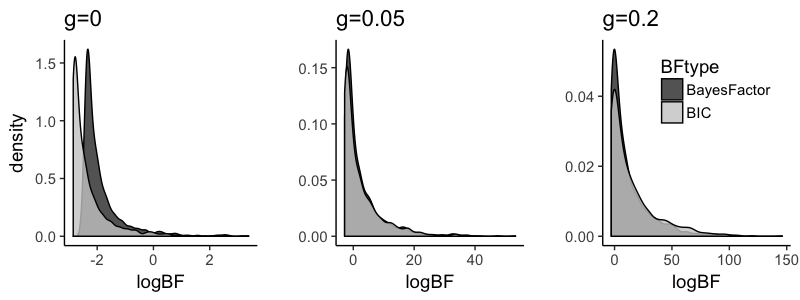}
  \caption{Kernel density plots of distributions of $\log BF$ for main effect $\alpha$, presented as a function of Bayes factor type (BIC versus BayesFactor) and effect size ($g=0, 0.05, 0.2$). }
  \label{fig:effectA}
\end{figure*}

As shown in Table \ref{tab:summary1}, the BIC approximation from Equation \ref{eq:BIC2} provides a similar distribution of Bayes factors compared to those computed from the BayesFactor package in R.  Figure \ref{fig:effectA} shows this pattern of results quite clearly, as the kernel density plots for the two different types of Bayes factors exhibit a large amount of overlap for the $g=0.05$ and $g=0.2$ conditions.  It is notable that the BIC approximation tended to underestimate the BayesFactor output in the $g=0$ case. However, as can be seen in the ``Consistency'' column of Table \ref{tab:summary1}, regardless of effect size conditions, the two different types of Bayes factors resulted in the same decision in a large proportion of simulations (at least 98.4\% of simulation trials).

\begin{figure*}
\centering
  \includegraphics[width=\textwidth]{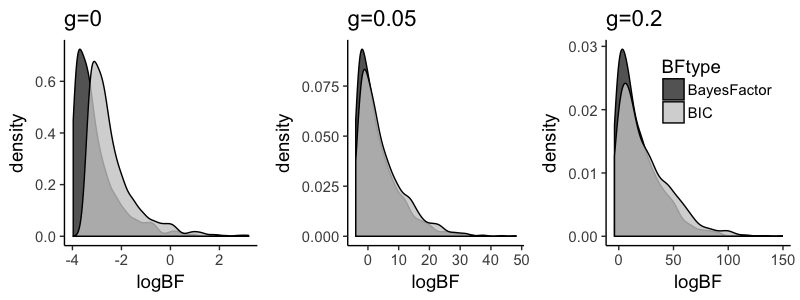}
  \caption{Kernel density plots of distributions of $\log BF$ for main effect $\tau$, presented as a function of Bayes factor type (BIC versus BayesFactor) and effect size ($g=0, 0.05, 0.2$). }
  \label{fig:effectB}
\end{figure*}

A similar picture emerges for the main effect $\tau$. As can be seen in Table \ref{tab:summary2} and Figure \ref{fig:effectB}, the BIC approximation and the BayesFactor outputs are largely consistent and result in mostly the same model choice decisions.  As with the results for main effect $\alpha$, there is some slight difference in the kernel density plots when simulating null effects (i.e., the condition $g=0$).  However, both methods chose the same model on at least 92.7\% of simulated trials, showing a good amount of consistency.

\begin{table*}
\caption{\label{tab:summary2}%
Summary of values for $\log BF$ for main effect $\tau$ with cell size $n=50$}
\begin{ruledtabular}
\begin{tabular}{cccccccc} 
  $g$ & $BF$ type & Min & $Q_1$ & Median & $Q_3$ & Max & Consistency\\
  \colrule
                    0 & BayesFactor & -3.97 & -3.68 & -3.31 & -2.70 & 2.25 & \\
                    & BIC & -3.40 & -3.10 & -2.70 & -2.05 & 3.19 & 0.980 \\[2mm]

                    0.05 & BayesFactor & -3.97 & -1.89 & 1.15 & 6.28 & 43.55 & \\
              & BIC & -3.40 & -1.05 & 2.28 & 8.07 & 48.17 & 0.927\\[2mm]

          0.2 & BayesFactor & -3.92 & 3.19 & 12.00 & 28.51 & 134.04 & \\
              & BIC & -3.34 & 6.03 & 17.15 & 36.58 & 149.39 & 0.952\\
\end{tabular}
\end{ruledtabular}
\end{table*}

\begin{figure*}
  \centering
  \includegraphics[width=\textwidth]{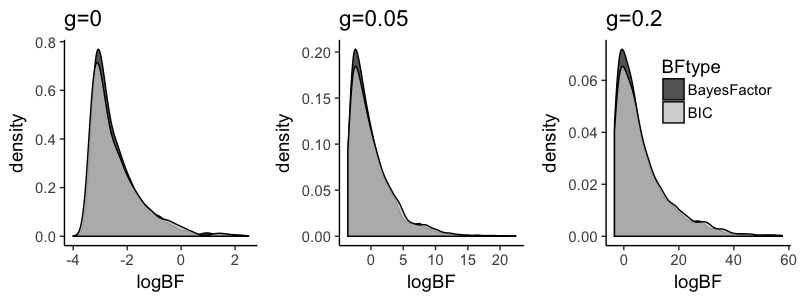}
  \caption{Kernel density plots of distributions of $\log BF$ for the interaction effect $\gamma$, presented as a function of Bayes factor type (BIC versus BayesFactor) and effect size ($g=0, 0.05, 0.2$). }
  \label{fig:effectAB}
\end{figure*}

Finally, we can see in Table \ref{tab:summary3} and Figure \ref{fig:effectAB} that the BIC approximation closely mirrors the output of the BayesFactor package for the interaction effect $\gamma$.  Indeed, the kernel density plots in Figure \ref{fig:effectAB} show considerable overlap between the distributions of BIC values and the distributions of BayesFactor outputs, and this picture is consistent across all three effect sizes ($g=0,0.05,0.2$).  As expected from this picture, both methods arrive at very similar model choices, picking the same model on at least 97\% of trials.

\begin{table*}
  \caption{\label{tab:summary3}
    Summary of values for $\log BF$ for interaction effect $\gamma$ with cell size $n=50$}
\begin{ruledtabular}	
\begin{tabular}{cccccccc} 
$g$ & $BF$ type & Min & $Q_1$ & Median & $Q_3$ & Max & Consistency\\
  \colrule
  0 & BayesFactor & -3.97 & -3.08 & -2.72 & -2.05 & 3.47 & \\
                    & BIC & -3.40 & -3.12 & -2.74 & -1.98 & 4.03 & 0.990 \\[2mm]

                    0.05 & BayesFactor & -3.57 & -2.30 & -0.96 & 1.24 & 20.64 & \\
              & BIC & -3.40 & -2.27 & -0.78 & 1.64 & 22.40 & 0.970\\[2mm]

          0.2 & BayesFactor & -3.37 & -0.40 & 3.26 & 9.57 & 54.17 & \\
              & BIC & -3.39 & -0.20 & 3.84 & 10.72 & 57.61 & 0.975\\
    
        \end{tabular}
\end{ruledtabular}
      \end{table*}

\section{Conclusion}
The BIC approximation given in Equation \ref{eq:BIC2} provides an easy-to-use estimate of Bayes factors for simple between-subject ANOVA and $t$ test designs.  It requires only minimal information, which makes it well-suited for using in a meta-analytic context.  In simulations, the estimates derived from Equation \ref{eq:BIC2} compare favorably to Bayes factors computed using existing software solutions with raw data.  Thus, the researcher can confidently add this BIC approximation to the ever-growing collection of Bayesian tools for scientific measurement.

\bibliography{references}

\begin{thebibliography}{}

\bibitem[Borota et~al., 2014]{borota2014}
Borota, D., Murray, E., Keceli, G., Chang, A., Watabe, J.~M., Ly, M., Toscano,
  J.~P., and Yassa, M.~A. (2014).
\newblock Post-study caffeine administration enhances memory consolidation in
  humans.
\newblock {\em Nature {N}euroscience}, 17(2):201--203.

\bibitem[Gigerenzer, 2004]{gigerenzer2004}
Gigerenzer, G. (2004).
\newblock Mindless statistics.
\newblock {\em The {J}ournal of {S}ocio-{E}conomics}, 33(5):587--606.

\bibitem[Hoekstra et~al., 2014]{hoekstra2014}
Hoekstra, R., Morey, R.~D., Rouder, J.~N., and Wagenmakers, E.-J. (2014).
\newblock Robust misinterpretation of confidence intervals.
\newblock {\em Psychonomic {B}ulletin {\&} {R}eview}, 21(5):1157--1164.

\bibitem[Jeffreys, 1961]{jeffreys1961}
Jeffreys, H. (1961).
\newblock {\em The {T}heory of {P}robability (3rd ed.)}.
\newblock Oxford University Press, Oxford, UK.

\bibitem[Lindley, 1957]{lindley1957}
Lindley, D.~V. (1957).
\newblock A statistical paradox.
\newblock {\em Biometrika}, 44(1-2):187--192.

\bibitem[Masson, 2011]{masson2011}
Masson, M. E.~J. (2011).
\newblock A tutorial on a practical {B}ayesian alternative to null-hypothesis
  significance testing.
\newblock {\em Behavior {R}esearch {M}ethods}, 43(3):679--690.

\bibitem[Morey and Rouder, 2011]{morey2011}
Morey, R.~D. and Rouder, J.~N. (2011).
\newblock Bayes factor approaches for testing interval null hypotheses.
\newblock {\em Psychological {M}ethods}, 16(4):406--419.

\bibitem[Morey and Rouder, 2015]{bayesfactor}
Morey, R.~D. and Rouder, J.~N. (2015).
\newblock {\em {BayesFactor}: {C}omputation of {B}ayes Factors for Common
  Designs}.
\newblock R package version 0.9.12-2.

\bibitem[Raftery, 1995]{raftery1995}
Raftery, A.~E. (1995).
\newblock Bayesian model selection in social research.
\newblock {\em Sociological {M}ethodology}, 25:111--163.

\bibitem[Rouder et~al., 2012]{rouder2012}
Rouder, J.~N., Morey, R.~D., Speckman, P.~L., and Province, J.~M. (2012).
\newblock Default {B}ayes factors for {ANOVA} designs.
\newblock {\em Journal of {M}athematical {P}sychology}, 56(5):356--374.

\bibitem[Rouder et~al., 2009]{rouder2009}
Rouder, J.~N., Speckman, P.~L., Sun, D., Morey, R.~D., and Iverson, G. (2009).
\newblock Bayesian $t$ tests for accepting and rejecting the null hypothesis.
\newblock {\em Psychonomic {B}ulletin {\&} {R}eview}, 16(2):225--237.

\bibitem[Sevos et~al., 2016]{sevos2016}
Sevos, J., Grosselin, A., Brouillet, D., Pellet, J., and Massoubre, C. (2016).
\newblock Is there any influence of variations in context on object-affordance
  effects in schizophrenia? {P}erception of property and goals of action.
\newblock {\em Frontiers in {P}sychology}, 7:1551.

\bibitem[Wagenmakers, 2007]{wagenmakers2007}
Wagenmakers, E.-J. (2007).
\newblock A practical solution to the pervasive problems of $p$ values.
\newblock {\em Psychonomic {B}ulletin {\&} {R}eview}, 14(5):779--804.

\bibitem[Wang, 2017]{wang2017}
Wang, M. (2017).
\newblock Mixtures of $g$-priors for analysis of variance models with a
  diverging number of parameters.
\newblock {\em Bayesian {A}nalysis}, 12(2):511--532.

\end{thebibliography}
\bibliographystyle{apalike}

\end{document}